\documentstyle[11pt]{article}

\begin{document}
\large

\def\lsim{\mathrel{\rlap{\lower3pt\hbox{\hskip0pt$\sim$}}
    \raise1pt\hbox{$<$}}}         %less than or approx. symbol
\def\gsim{\mathrel{\rlap{\lower4pt\hbox{\hskip1pt$\sim$}}
    \raise1pt\hbox{$>$}}}         %greater than or approx. symbol
\def\dblint{\mathop{\rlap{\hbox{$\displaystyle\!\int\!\!\!\!\!\int$}}
    \hbox{$\bigcirc$}}}
\def\ut#1{$\underline{\smash{\vphantom{y}\hbox{#1}}}$}

\newcommand{\beq}{\begin{equation}}
\newcommand{\eeq}{\end{equation}}
\newcommand{\dem}{\Delta M_{\mbox{B-M}}}
\newcommand{\dega}{\Delta \Gamma_{\mbox{B-M}}}

\newcommand{\ind}[1]{_{\begin{small}\mbox{#1}\end{small}}}
\newcommand{\WA}{{\em WA}}
\newcommand{\SM}{Standard Model }
\newcommand{\QCD}{{\em QCD }}
\newcommand{\KM}{{\em KM }}
\newcommand{\hscale}{\mu\ind{hadr}}
\newcommand{\sG}{i\sigma G}

\newcommand{\MS}{\overline{\mbox{MS}}}
\newcommand{\pole}{\mbox{pole}}
\newcommand{\aver}[1]{\langle #1\rangle}

\newcommand{\appa}{\mbox{\ae}}
\newcommand{\CP}{{\em CP } }
\newcommand{\fy}{\varphi}
\newcommand{\hi}{\chi}
\newcommand{\al}{\alpha}
\newcommand{\as}{\alpha_s}
\newcommand{\gf}{\gamma_5}
\newcommand{\gm}{\gamma_\mu}
\newcommand{\gn}{\gamma_\nu}
\newcommand{\be}{\beta}
\newcommand{\ga}{\gamma}
\newcommand{\de}{\delta}
\renewcommand{\Im}{\mbox{Im}\,}
\renewcommand{\Re}{\mbox{Re}\,}
\newcommand{\GeV}{\,\mbox{GeV}}
\newcommand{\MeV}{\,\mbox{MeV}}
\newcommand{\matel}[3]{\langle #1|#2|#3\rangle}
\newcommand{\state}[1]{|#1\rangle}
\newcommand{\ra}{\rightarrow}
\newcommand{\ve}[1]{\vec{\bf #1}}

\newcommand{\rhs}{{\em rhs}}
\newcommand{\pp}{\langle \ve{p}^2 \rangle}

\newcommand{\BR}{\,\mbox{BR}}
\newcommand{\La}{\overline{\Lambda}}

\vspace*{.7cm}
\begin{flushright}
\large{
CERN-TH.7063/93\\
UND-HEP-93-BIG\hspace*{0.1em}05}\\
hep-ph/9311243\\
revised version
\end{flushright}
\vspace{1.2cm}
\begin{center} \LARGE {$D_s$ Lifetime,  $m_b$, $m_c$ and $|V_{cb}|$
in the Heavy Quark Expansion }
\end{center}
\vspace*{.4cm}
\begin{center} \Large
I.I.Bigi\\
{\normalsize {\it Theoretical Physics Division, CERN, CH-1211
Geneva 23, Switzerland \footnote{during the academic year
1993/94}}\\
and\\
{\it Dept.of Physics,
University of Notre Dame du
Lac, Notre Dame, IN 46556 \footnote{permanent address} } \\
{\it e-mail address: IBIGI@CERNVM, BIGI@UNDHEP}}
\vspace{.4cm}
\\
\Large
N.G.Uraltsev
\\{\normalsize{\it
Theoretical Physics Division, CERN, CH-1211
Geneva 23, Switzerland}
\\and\\
{\it St.Petersburg Nuclear Physics Institute,
Gatchina, St.Petersburg 188350, Russia $^2$
\\
e-mail
address:  URALTSEV@UNDHEP ;  URALTSEV@LNPI.SPB.SU}}
\end{center}
\thispagestyle{empty} \vspace{.4cm}

\centerline{\Large\bf Abstract}
\vspace{.4cm}
We present some straightforward applications of the QCD
heavy quark expansion, stated
in previous papers [1-3],
to the inclusive widths of heavy flavour hadrons.
We address the question of the $D_s$
lifetime and argue that -- barring Weak Annihilation (WA) --
$\tau (D_s)$ is expected to exceed $\tau (D^0)$ by several percent;
on the other hand WA could
provide a difference of up to $10\div20\%$ of
{\it any} sign. We extract $m_c$,
$m_b$ and $|V_{cb}|$ from $\Gamma\ind{SL}(D^+)$ and $\Gamma\ind{SL}(B)$.
The
values of the quark masses are somewhat higher, but
compatible with estimates from QCD sum rules; we obtain
$|V_{cb}|\simeq 0.043$ for $\tau (B)=1.4$ psec and
$BR_{SL}(B)=10.5$\% .
We discuss the associated uncertainties in the $1/m_Q$
expansion as well as some consequences for other electroweak decays.

\newpage
\large
\addtocounter{footnote}{-2}

Our theoretical understanding of both exclusive and inclusive
heavy flavour decays has improved considerably over the last few
years. This progress has been driven largely by treatments that involve
expanding weak transition amplitudes in terms of $1/m_Q$, $m_Q$ being
the mass of the heavy flavour quark, i.e.
$m_Q=m\ind{beauty},\; m\ind{charm}$. In this note we will analyze some
phenomenological consequences that such a general
treatment has for inclusive decays of beauty and charm hadrons.

The paper will be organized as follows : in Sect.1 we
recapitulate briefly the salient features of our method which in Sect.2
is then applied to the lifetime of $D_s$ mesons; in Sect.3 we
extract the size of $m_c$ from the semileptonic $D$ width and infer
the value of $m_b$; the value of $|V_{cb}|$ is then determined from the
semileptonic $B$ width. In Sect.4 we present our conclusions.

\section {General Method}

On very general grounds one expects that the spectator ansatz
correctly describes inclusive heavy flavour decays for
asymptotically large values of $m_Q$; yet at finite $m_Q$
there arise nonperturbative corrections to transition rates
that among other things generate different lifetimes for the various
types of hadrons $H_Q$ carrying the flavour $Q$. It was shown in
refs.~[1-5] how these preasymptotic effects can be
incorporated in a self-consistent and systematic way;
this will be exemplified now for the case of charm
lifetimes.

The width for the weak decay of a charmed hadron into an inclusive final
state $f$ is obtained from the transition operator that has
been expanded
into a series of local operators \footnote{This expansion is based on a
sufficiently large energy release. It is not applicable for
beauty decays into
$\tau$ leptons, whose partial width is to be calculated explicitly. It
is, however, the relevant approach to other annihilation-induced
semileptonic widths \cite{BU2}.}:
$$\hat{\Gamma}=
\frac{G_F^2m_c^5}{192\pi^3} |KM|^2 \{c_0(f) \bar{c}c +
\frac{c_2(f)}{m_c^2}\bar{c} i\sigma G c +
\Sigma_i\frac{c_3(f)}{m_c^3} (\bar{c}
\Gamma _i q)(\bar{q} \Gamma _ic) + {\cal O} (1/m_c^4)\}  \eqno(1)$$
where the dimensionless coefficients $c_i(f)$ depend on
the parton level characteristics of $f$ and on the ratios of
masses of the final state quarks to the
mass of the $c$ quark; $KM$ denotes the appropriate product of the weak
mixing angles.

The operators that appear on the right hand side of eq.(1) represent
a rather universal cast, e.g. they control semileptonic as well as
nonleptonic transitions and also distributions like semileptonic
spectra:

$\bullet$ the scalar density $\bar{c} c$ describing the `quasifree' decay of
the $c$ quark;

$\bullet$ the chromomagnetic dipole operator $\bar c\, \sG\, c$;

$\bullet$ the four-fermion operator $(\bar c \Gamma q)(\bar q\Gamma c)$.
It actually contains terms differing in both the flavour of the
light quark $q$ and the particular Lorentz structure of the
$\gamma$-matrices $\Gamma _i$.

The widths for $D_s$, $D^0$, $D^+$ and $\Lambda _c$
decays are then given by the diagonal matrix
elements of the operator $\hat{\Gamma}$ from eq.(1) between the
corresponding states $D_s$, $D^0$, $D^+$ and $\Lambda _c$. These
matrix elements depend on long distance bound state dynamics; thus they
cannot be evaluated perturbatively. Yet this aspect of the strong
interaction cannot be ignored, for even the matrix element of the
`quasifree' operator $\bar c c$ contains nonperturbative effects,
giving rise
to some nontrivial width splitting:
$$\matel{H_c}{\bar{c}c}{H_c}=
2M_{H_c}+\frac{1}{4m_c^2}\matel{H_c}{\bar{c}i\sigma G c}{H_c}-
\frac{1}{2m_c^2}\matel{H_c}{\bar{c}(i\ve D)^2 c}{H_c}
+{\cal O}(1/m_c^3) \eqno(2)$$
with $D_{\mu}$ denoting the covariant derivative; a relativistic
normalization has been employed here.
In addition to the chromomagnetic operator another
operator appears now, namely $\bar c(i\ve D)^2c$ which
describes the kinetic energy of the charm quark in the gluon background
field.

The size of the matrix element of the chromomagnetic operator between
the heavy flavour mesons can be extracted from the hyperfine spin splitting
between the pseudoscalar and vector states:
$$ \frac{1}{2M_c}\matel{D}{\bar{c} \sG c}{D} \simeq
\frac{3}{2}\cdot(M_{D^*}^2-M_D^2) \;\;.\eqno(3)$$
For baryons it vanishes:
$$\matel{\Lambda_c}{\bar{c}i\sigma G c}{\Lambda_c}\simeq 0\; .
\eqno(4)$$
The following general conclusions can be drawn from this approach:

\noindent $\bullet$ The leading nonperturbative
corrections arise on the $1/m_c^2$ level.

\noindent $\bullet$ There are two distinct sources of
such $1/m_c^2$ corrections:
(i) The matrix elements of the $d=5$
operator $\bar{c} \sG c$ that appear in eq.(1); its weight depends on
which particular channel is considered.
(ii) The expansion of the matrix elements of
the `quasifree' operator $\bar{c} c$, see eq.(2)
that affects uniformly all
decay modes of a particular hadron.

\noindent $\bullet$ These corrections are flavour independent: they
do not depend explicitly on the flavour of the spectator
antiquark in the meson; they do, however,
differentiate between mesons and baryons.

\noindent $\bullet$ Order $1/m_Q^3$ corrections are produced
by the four-fermion operators
$(\bar c \Gamma _iq)(\bar q\Gamma _ic)$
describing the nonspectator effects of `Pauli Interference' (PI)
and `Weak Annihilation' (WA) or `Weak Scattering' (WS) in baryons.
These flavour-dependant operators
generate different lifetimes even among the members of the same
isomultiplet\footnote{Similar effects leading to
the width splitting of the two $D^0$ eigenstates are described by the
corresponding four fermion operator of the form $(\bar{c}u)^2$. Its effect,
however, is strongly suppressed and does not shift the central
value of the width; we ignore it in our analysis.}.

The analysis of $1/m_c^2$ [1-4] and
$1/m_c^3$ \cite{VSlog,BU,BU2} corrections shows
that their size is in general quite large in
charm decays. For example the chromomagnetic operator produces corrections
$\sim [3(M^2_{D^*}-M^2_D)]/(2m^2_c)\simeq 0.4$.
Higher order corrections can thus be expected to be still significant
and we have to be satisfied with a typically
semiquantitative analysis. For example the theoretical predictions
$\tau (D^+)/\tau (D^0) \sim 2$,
BR$\ind{SL}(D^+)\sim 16\%$ and BR$\ind{SL}(D^0)\sim 8\%$
have to be seen as in agreement with the
experimental findings within the uncertainties of such a treatment.

\section{$D_s$ vs. $D^0$ Lifetimes}

The lifetimes of $D^0$ and $D_s$ mesons could {\em a priori} differ
substantially from each other due to $SU(3)\ind{Fl}$ breaking and in
particular due to a different
weight of WA in $\Gamma (D^0)$ and
$\Gamma (D_s)$. A previous analysis \cite{BU}
suggested WA to be smallish already in charm meson decays;
furthermore $1/m_Q$ expansions naturally lead to the prediction
of $SU(3)\ind{Fl}$ breaking to be small in heavy flavour
decays, namely of the order of several percent only \cite{BUV}.
Thus one predicts
$$\tau (D_s)\simeq \tau(D^0) \eqno(5)$$
to first approximation, in agreement with recent E687 data:
$$ \frac {\tau (D_s)}{\tau (D^0)}= 1.13 \pm 0.05 \eqno(6)$$
A rather delicate analysis is required to go beyond the
semi-quantitative prediction of eq.(5) to see whether, indeed,
the $D_s$ lifetime should be slightly longer than the
$D^0$ lifetime, as suggested by present data, and by how much.

There are two main sources for lifetime differences among
charm mesons, namely explicitly flavour-dependent $1/m_c^3$
terms and corrections of order $m_s\hscale/m_c^2$ due
to $SU(3)\ind{Fl}$ breaking in the `flavour-independent'
$1/m_c^2$ contributions.

Attempting to estimate relative corrections of a few percent in
nonleptonic $D$ decays is a bold enough enterprise and we
will ignore processes contributing less than 1\% to the
total width. In that category are {\em e.g.} doubly Cabibbo suppressed
transitions $c\ra d\bar s u$ with a relative rate
$\propto \tan ^4\theta _c \simeq 3\cdot 10^{-3}$ as well as
Penguin driven processes; the latter are suppressed
by either tiny mixing angles with the b quark or by the small
mass of the s quark.

Numerically large effects can arise only from Cabibbo-allowed
channels: WA is present in nonleptonic $D^0$ decays, and in
semileptonic as well as nonleptonic $D_s$ decays; it also
drives $D_s\ra \tau \nu$. Cabibbo-suppressed modes
$c\ra s\bar su,\; d\bar du$ can produce corrections to the total
width of a few percent: WA affects $D^0$, $D^+$ and $D_s$ decays; in
addition to $D^+$ PI intervenes also in $D_s$ modes.

Since the impact of WA as compared to PI
is reduced in meson decays \cite{BU2,VSlog,BU,BSold},
it is natural to compare $\tau (D_s)$ with
$\tau (D^0)$ rather than with $\tau (D^+)$. There are four
distinct sources for a difference in $\Gamma (D_s)$
vs. $\Gamma (D^0)$ exceeding the 1\% level:\\
{\bf (a)} The decay
$D_s\ra \tau \nu$. {\bf (b)} PI in those Cabibbo suppressed $D_s$ decays that
are driven by the quark level transition $c\ra s \bar s d$. {\bf (c)} The
effects of $SU(3)$ breaking on the matrix elements of the chromomagnetic
and kinetic energy operators. {\bf (d)} WA in nonleptonic $D^0$ and in
nonleptonic as well as semileptonic $D_s$ decays.

While the corrections listed under (a) and (d)
have been discussed extensively in the literature,
the corrections referred to under (b) and especially (c) on the other hand
are much less familiar. \vspace{.15cm}

{\bf (a)} The width for the decay $D_s\ra \tau\nu_\tau$ is completely
determined in terms of the axial decay constant for the $D_s$ meson:
$$\Gamma(D_s\ra \tau\nu_\tau) \simeq \frac {G_F^2 m_\tau^2f_{D_s}^2 M_{D_s}}
{8\pi} |V_{cs}|^2 (1-m_\tau^2/M_{D_s}^2)^2  \;\;\;.\eqno(7)$$
For $f_{D_s}\simeq 210\MeV$ one gets numerically
$$\Gamma(D_s\ra \tau\nu_\tau) \simeq .03\, \Gamma(D^0)\;\;\;.\eqno(8)$$
This effect necessarily reduces $\tau (D_s)$ relative to $\tau (D^0)$.
\vspace{.15cm}

{\bf (b)} PI in $D_s$ appears in the $c\ra
s\bar{s} u$ channel.
Its weight is expressed in terms of the matrix elements
of the four-fermion operators
$$\matel{B_s}{(\bar c_L\gm s_L)(\bar s_L \gm
c_L)}{B_s}\;\;\;,\;\;\; \matel{B_s}{(\bar c_L\gm \frac{\lambda^a}{2} s_L)
(\bar s_L \gm \frac{\lambda^a}{2} c_L)}{B_s}  \eqno(9)$$
with known coefficients that are computed perturbatively,
see refs.\cite{VSlog,BU,BU2}; the hybrid renormalization \cite{VSlog} of
these operators down from the scale $m_c^2$ has to be included. The
most reliable way to estimate the effect of PI, we believe, is to
relate it to the observed difference in the $D^+$ and $D^0$ widths.
Both PI and WA contribute in reality to this difference; yet
according to a detailed analysis \cite{BU,BSold,BU2} PI
is the dominant effect in mesons (see also the discussion below).
It is also
worth noting that the size of the PI
correction in $D^+$ as
estimated theoretically reproduces the observed
width splitting for reasonable values of $f_D$ -- provided the hybrid
renormalization of the operators involved is taken into account.

It is easy to see that the structure of the operators responsible for PI
in $D_s$ is exactly the same as in $D^+$ if one replaces
the $d$ quark by the
$s$ quark and adds the extra factor $\tan^2\theta_c$; it is then
{\em destructive} as well. From the assumed dominance of PI in the
$D^+$-$D^0$ lifetime difference one thus arrives at:
$$\de \Gamma\ind{int}(D_s)\simeq S\cdot \tan^2(\theta_c)
(\Gamma(D^+)-\Gamma(D^0)) \simeq -S \cdot 0.03 \Gamma(D^0) \eqno(10) $$
where
the factor $S$ has been introduced to allow for $SU(3)$ violation in the
relevant matrix elements of the four-fermion operators
\footnote{$SU(3)_{Fl}$
violation also affects the coefficients of the operators that depend on the
mass of the quarks in the final state; those corrections, however,
are of the
order of $(m_s/m_c)^2$ whereas the factor $S$ is linear in
$m_s/\mu_{\mbox{hadr}}$, and we neglect the former.}.

 The factor $S$ is expected to exceed unity somewhat;
in the factorization approximation it is given by the ratio
$(f_{D_s}/f_D)^2$.
Various estimates yield the range $S=1 \div 1.7$; to be more specific
we adopt $S=1.4$. Then we conclude that
$$ \delta \Gamma \ind{int}(D_s) \sim - 0.04\Gamma(D^0) \;\;\;.\eqno(11)$$
\vspace{.15cm}

{\bf (c)} As already stated in Sect.1 the $1/m_c^2$ corrections are given
by the appropriate expectation values
of two dimension five operators. As far
as the chromomagnetic operator is concerned one has the general
expressions:
$$ \frac{1}{2M_c}\matel{D^0}{\bar{c} \sG c}{D^0} \simeq
\frac{3}{2}\cdot(M_{D^{0*}}^2-M_{D^0}^2) \;\;.\eqno(12a)$$
$$ \frac{1}{2M_c}\matel{D_s}{\bar{c} \sG c}{D_s} \simeq
\frac{3}{2}\cdot(M_{D_s^*}^2-M_{D_s}^2) \;\;.\eqno(12b)$$
Since the measured values for $D$-$D^*$ and for the $D_s$-$D_s^*$ mass
splittings are almost identical, the chromomagnetic operator
cannot be expected to induce an appreciable difference between $\tau
(D_s)$ and $\tau (D^0)$.

The observation that the hyperfine splitting is
largely independent of the flavour of the spectator can be understood in
the following intuitive picture (see ref.~\cite{RW}): using a simple
constituent description one finds that the chromomagnetic field strength is
proportional both to the chromomagnetic dipole moment of the spectator (or
more generally, of the light degrees of freedom in the $D$ meson) and to the
wavefunction density at the origin, $1/\aver{r}^3$.  The former is
most naturally expected to decrease when the (current)
mass of the spectator
quark increases,
whereas the latter is always assumed to increase when going from
non-strange to strange particles. Such a behaviour
can explicitly be demonstrated
at least in the limit when the spectator becomes
heavy enough as well
\footnote{The situation with the chromomagnetic dipole
moment could in principle have been more
complicated if the nonspectator
(gluon) light degrees of freedom tended to form a ground state with a
nonzero spin inside a meson;
then for an arbitrarily weak interaction of the
spectator's spin the chromomagnetic term would be still finite. This
abstract possibility however contradicts the
spectrum of states in heavy
quarkonia -- it would produce an
additional hyperfine splitting pattern due
to interaction of the heavy spins with the
spin of the light degrees of
freedom.}.
These two effects may thus offset each other.
The conclusion about the equal strenght of the hyperfine splitting
can and of course must be tested in B mesons where the mass formulae are
more reliable due to the larger $b$ quark mass.

The second operator that generates $1/m_c^2$ corrections is the kinetic
operator $\bar{c}(i\ve D)^2 c$ which describes a nonrelativistic (``Fermi'')
motion of the charm quark. As mentioned above one expects the spatial
wavefunction to be more concentrated around the origin for $D_s$ than for
$D$ mesons. This in turn implies that the mean value of $\ve{p}^2$ is
expected to be larger for $D_s$ than for $D$ mesons; in other words the
charm quark undergoes more Fermi motion as a constituent of $D_s$ than of
$D$ mesons.  Correspondingly the lifetime of the charm quark is prolonged by
time dilation to a higher degree inside $D_s$ than inside $D$ mesons. Eq.(2)
makes this connection quite explicit \cite{BBSUV,BSUV}:  the factor $1-\pp
/2m_c^2$ appearing in the expression for the transition operator is actually
nothing but the mean value of the Lorentz factor $\sqrt{1-v^2}$ that
suppresses the decay probablity of a particle in a moving frame.

The qualitative trend of this effect is thus quite transparent. However its
numerical size is not, at least not yet. One can expect future
progress in QCD sum rules and/or simulations of QCD on the lattice to
determine the appropriate matrix elements
$\matel{H_c}{\bar{c}(i\ve D)^2c}{H_c}$.
Yet for the problem at hand, namely the $D_s$-$D^0$ lifetime
difference one can in principle extract the relevant matrix element from the
measured values of meson masses in the charm and beauty sector according to
the following prescription~\cite{BUV1}:
$$
\frac{1}{2M_D}(\matel{D_s}{\bar{c}(i\ve D)^2 c}{D_s} -
\matel{D}{\bar{c}(i\ve D)^2 c}{D}) \simeq $$
$$\simeq \frac{2m_bm_c}{m_b-m_c} \{[\aver{M_{D_s}}-\aver{M_{D}}] -
[\aver{M_{B_s}}-\aver{M_B}]\}\;\;\; \eqno(13)$$
where $\aver{M_{D,D_s,B,B_s}}$
denote the `spin averaged' meson masses: e.g. for D mesons
$$\aver{M}_D=\frac{M_D+3M_{D^*}}{4}$$
and likewise for the other mesons. Accordingly one finds
$$\frac{\Delta\Gamma\ind{Fermi}(D_s)}{\bar{\Gamma}}\simeq
- \frac{m_b}{m_c(m_b-m_c)} \{[\aver{M_{D_s}}-\aver{M_{D}}]
- [\aver{M_{B_s}}-\aver{M_B}]\}\;\;\;. \eqno(14)$$

A $10\MeV$ shift in any of the `spin averaged' mass terms
$\aver{M}$ in eq.(14) corresponds numerically
to the kinetic energy operator generating
approximately a 1\% change in the ratio
$\tau (D_s)/\tau (D^0)$.  The meson
masses have been measured with an accuracy of 2 MeV or better,
which is
sufficient for our analysis -- with the exception of the $B_s$-$B^*_s$
sector.  Very recent LEP/CDF
data \cite{Bsmass} indicate that the $SU(3)$ mass
splitting in the beauty sector is practically the same
as in the charm sector, namely
$M_{B_s}-M_{B^0} \simeq (94.5\pm 4.6\MeV)$
vs. $M_{D_s}-M_{D} \simeq (99.5\pm 0.7\MeV)$.
Nothing is known experimentally
about the $B_s^*$ mass; on the other hand retaining only the leading
non-trivial term in a heavy quark expansion, one would conclude that the
hyperfine splitting in the $B_s$-$B_s^*$ system be the same as in the
$B_d$-$B_d^*$ --  simply because the analogous equality holds for charm
mesons. Combining the preliminary measurement of the $B_s$ mass with these
{\em theoretical} expectations about the $B_s^*$ mass would lead to the
rather surprising result that the mean momentum of the heavy quark is
practically the same in strange and non-strange
heavy-flavour mesons; the
Fermi motion of the charm quark could then cause a difference in
$\tau (D_s)$ vs. $\tau (D^0)$ of at most $1\%$.

This conclusion can be confronted with a general expectation: it is natural
to expect the value of $\pp$ to be at least of order of $(400\MeV)^2$ even
in non-strange heavy mesons\footnote{A recent QCD sum rule estimate yielded
about $0.5\div 0.6\GeV^2$ \cite{VLB}.}; this by itself would suppress the
width in charm by about $3\%$ and increase the mass of the meson by
$50\MeV$.  Assuming that $SU(3)$ violation increases $\pp$ by a moderate
factor $1.5$ in the strange system one would then expect at least a $2\%$
decrease in the width of $D_s$.

We think that this oversimplified yet transparent
line of reasoning cannot be ruled
out yet; it would seem quite premature to conclude that the
Fermi motion plays a negligible role in the $D_s$-$D^0$ lifetime
difference. Namely the chromomagnetic field in $D_s$ and $D$ may not coinside
so closely to predict the $B_S^*$ mass with the necessary precision.
For one has to allow for sizeable corrections to the mass formulae
stated above where only the leading terms in $1/m_b$ and
$1/m_c$ have been retained since $m_c$ is not much larger than
typical hadronic scales. There is no general reason to expect
that the mass splitting formulae hold to better than 30\%
accuracy. Taking into account that the size of the hyperfine splitting in
charm mesons constitutes about $140\MeV$, such an uncertainty translates
into a 3-4\% shift in the lifetime ratio, see eq.(14).
Therefore we conclude:  ($\alpha$) The Fermi motion
of the charm quark inside the meson may well prolong $\tau (D_s)$ by
typically a few percent relative to $\tau (D)$.  ($\beta$) A better
determination of the chromomagnetic field can be obtained from the $B$
system. Therefore a more definite predictions can be made once the mass of
the vector meson $B_s^*$ has been measured and the experimental uncertainty
on $M(B_s)$ has been decreased.

One further comment is in order here. The discussion given
above was concerned with
nonleading corrections to the mass formulae for charm particles
that could be sizeable because of the moderate value of $m_c$.
Could there be analogous corrections to the expansion of the {\em
widths} of charm particles?  In particular there could be a sizeable
deviation from the relation $\matel{D_s}{\bar c \sG c}{D_s} \simeq
\matel{D}{\bar c \sG c}{D}$; this would
show up as a violation of the prediction
$M_{B_s^*}-M_{B_s}\simeq M_{B^*}-M_{B}$ that is suggested by a
simple
extrapolation from charm. This
deviation would reflect the weight of nonleading
corrections in the charm system and
could {\em a priori} have a significant impact
on the width; for the chromomagnetic
interaction seems to be more important numerically
than the kinetic energy \cite{BSUV}. However we do not expect
this to amount to an important effect. For the chromomagnetic operator
appears twice, namely in the expansion of the transition operator $\hat
\Gamma$, eq.(1), and in the `quasi-free' operator $\bar cc$; there is
an almost complete
cancellation between the chromomagnetic contributions from the
nonleptonic and the two semileptonic channels if one
uses for the coefficient $c_2$ in eq.(1) the expression obtained in
ref.~\cite{BUV}. Even allowing for a $30\%$ uncertainty in the coefficient
$c_2$ the shift cannot exceed  $1\%$ for each $30\MeV$ in the
hyperfine splitting of $D$ or $D_s$.  (A more
accurate prediction had to include the next to leading {\em perturbative}
corrections in the coefficient $c_2$; computing them represents a task that
is straightforward in principle, although tedious in practice.)
\vspace*{.15cm}

{\bf (d)} The numerical impact of WA on charm meson lifetimes is
the most obscure theoretical item in the analysis.
The task becomes even harder for our present analysis addressing a
difference in the WA contribution to $\tau (D_s)$ and to
$\tau (D^0)$.
The uncertainty centers mainly on the question of how much the WA
amplitude suffers from helicity suppression. In the valence quark
description the answer is easily given to lowest order: the WA
rate is suppressed by
the ratio $m^2_q/m_c^2$ where $m_q$ denotes the largest quark mass in
the final state. For a proper QCD treatment one has to use current
rather than the larger constituent quark masses, at least for the
$1/m_c^3$ corrections. That would mean that WA is
 negligible in $D^0$ decays where the appropriate factor reads
$(m_s/m_c)^2\lsim .01$ and {\em a fortiori} in $D_s$ decays where only
non-strange quarks are present in the final state.  (Semi-)Hard gluon
radiation cannot circumvent this suppression \cite{BU}. For such gluon
corrections -- when properly accounted for -- drive the hybrid
renormalization of the corresponding four fermion operators which however
preserves their Lorentz structure and therefore does not eliminate the
helicity
suppression, at least in the leading {\em log} approximation. A helicity
allowed amplitude can be induced only at the subleading $\al_s(m_c^2)/\pi$
level and is thus expected to be numerically insignificant.

On the other hand nonperturbative dynamics can quite naturally vitiate
helicity suppression and thus provide the dominant source of WA. These
nonperturbative effects enter through nonfactorizable contributions
to the hadronic matrix elements. This has been analyzed in considerable
technical detail in ref.\cite{BU2}. The expression for the weak
annihilation operator in $D_s$ as well as its renormalization can be
easily obtained from the general expression for the case of $D^0$ or
$B^0$ (see refs.~\cite{BU,BU2}) by interchanging the colour factors $c_1$ and
$c_2$:
$$ \hat{\Gamma}\ind{ann} \simeq -\frac{G_F^2|V_{cb}|^2}{6\pi}
(p^2\de_{\mu\nu}-p_\mu p_\nu) (\,a\ind{sing}
(\bar c_L \gm q_L)\,(\bar q_L \gm c_L)
\, + \, a\ind{oct} (\bar c_L \gm \frac{\lambda^a}{2} q_L)\,(\bar q_L \gm
\frac{\lambda^a} {2}c_L)\:) \;;$$
$$a\ind{sing}=(3c_2^2+2c_1c_2)\appa^{9/2}+\frac{1}{3}c_1^2-\frac{1}{9}
(\appa^{9/2}-1)(3c_2^2+2c_1c_2)\;\;\;,$$
$$a\ind{oct}=2c_1^2-(\appa^{9/2}-1)(2c_2^2+\frac{4}{3}c_1c_2)\;\;\;,$$
$$c_1\simeq \frac{c_+ + c_-}{2}\;\;,\;\; c_2\simeq
\frac{c_+-c_-}{2}\;\;,\;\;q=d\;\;\;\mbox{for $D^0$ decays}\;\;,$$
$$c_1\simeq \frac{c_+-c_-}{2}\;\;,\;\; c_2\simeq
\frac{c_++c_-}{2}\;\;,\;\;q=s\;\;\mbox{for $D_s$ decays}\;\;,$$
$$
\appa=[\frac{\al_s(\mu^2)}{\al_s(m_c^2)}]^{1/b}\;\;\;,\;\;\;
b=11-\frac{2}{3}n_f=9\;\;\;.\eqno(15)$$
For each of the two semileptonic channels in $D_s$ one additionally has
$$a\ind{sing}=\frac{8}{9}\appa^{9/2}+\frac{1}{9}
\;\;\;,\;\;\;
a\ind{oct}=-\frac{2}{3}(\appa^{9/2}-1)\;\;\;.\eqno(15a)$$
Numerically in $D_s$ the overall singlet coefficient $a_s$ is about
$9$ and
the octet one is $-2.3$ whereas in $D^0$ they are $-.32$ and $3.2$
respectively.

The reason behind $a\ind{sing}$ being so much larger for $D_s$ than for
$D^0$ mesons is that there is a colour-allowed WA contribution to
$D_s$ decays while WA is colour-suppressed in $D^0$ decays.
This colour-allowed contribution is obviously factorizable, however the
factorizable part practically vanishes due to helicity suppression.
Appreciable effects can then come only from nonfactorizable
contributions or from ${\cal O}(\alpha _S(m_c^2)/\pi )$
corrections to the leading $\log$ approximation. Naive colour counting
rules suggest that nonfactorizable parts in the matrix elements
of colour singlet operators are $1/N_c$ suppressed as compared
to those of colour octet operators. This line of reasoning is
at best semi-quantitative, but if one adopts it one would conclude
that the weight of WA is similar in inclusive $D_s$ and
$D^0$ decays. As already stated nonleading perturbative
corrections are capable of producing helicity unsuppressed
contributions even to factorizable matrix elements; yet also
they are colour-suppressed.

It was shown in ref.~\cite{BU2} that a detailed experimental
study of the semileptonic width and the lepton
spectrum, in particular in
the endpoint region, in $D^0$ vs. $D^+$ and/or
in $B^0$ vs. $B^+$ decays
would allow us
to extract size of the matrix elements that control the
weight of WA in
all inclusive $B$ and $D$ decays.
Since such data are not (yet) available we can at present draw only a
qualitative conclusion:  WA is not expected
to affect the total lifetimes of
$D^0$ and $D_s$ mesons by more than
$10 \div 20 \%$, see refs.~\cite{BSold,BRAUN}.
Furthermore WA does not necessarily enhance $\tau (D_s)/\tau (D^0)$:
due to its interference with the spectator reaction it could even
reduce it!
\vspace*{.15cm}

To summarize our findings on the $D_s-D^0$ lifetime ratio:
$SU(3)\ind{Fl}$ breaking in the leading nonperturbative corrections
of order $1/m_c^2$ can -- due to `time dilatation' -- increase
$\tau (D_s)$ by $3\div 5\%$. On the $1/m_c^3$ level there arise
three additional effects. Destructive interference in Cabibbo
suppressed $D_s$ decays increases $\tau (D_s)$ again by
$3\div 5\%$ whereas the mode $D_s\ra \tau \nu$ decreases it
by $3\%$. These three phenomena together lead to
$\tau (D_s)/\tau (D^0)\simeq 1.0\div 1.07$. Any difference
over and above that has to be attributed to WA. Taking these
numbers at face value one can interprete the recent
measurement \cite{Ds}
of the $D_s$ lifetime in turn as more or less direct
evidence for WA to contribute not more than 10-20 \% of the
lifetime ratio between charm mesons. As expected
\cite{VSlog,BU}, it does not constitute the major effect
there. Finally the predictions just stated can be refined
by future more accurate measurements, namely

$\bullet$ of the difference in the
semileptonic spectra of charged and neutral
mesons in the charm and in the beauty sector to extract the
size of the matrix elements controlling the weight of WA;

$\bullet$ of $M(\Lambda _b)$, $M(B_s)$ and $M(B_s^*)$ to
better than 10 MeV to determine the expectation values of the
kinetic energy operator.

\section{Heavy Quark Masses and $|V_{cb}|$ -- phenomenological approach}

Conventional wisdom has it -- based on considering the simplest
perturbative diagrams -- that semileptonic decays are easier to treat
theoretically since they are less affected by the strong interactions.
Our analysis of non-perturbative corrections
in inclusive heavy flavour decays [1-5]
offers additional evidence in support of this conviction. For there
are smaller and fewer nonperturbative corrections in semileptonic
than in nonleptonic decays of heavy flavour mesons.
Even the first, $\cal{O}(\as)$ perturbative corrections may appear to be quite
large in nonleptonic $b$ decays owing to a relatively large mass
of the charm quark in the final state \footnote{We are grateful to V.Braun
for presenting explicit arguments in favour of this option and to M.Shifman
for a discussion of this problem.}.
%\footnote{From that point of view
%the semileptonic decays of $\Lambda_c$ would seem to provide an ideal lab.
%However at present there are no accurate data available on the semileptonic
%width of $\Lambda_c$; furthermore the flavour dependent effects possess
%there a more complex structure.}
The semileptonic widths are then best suited to extract fundamental
parameters like quark masses and KM angles.

The semileptonic width of the $D$ meson is given by the expansion
$$\Gamma(D\ra l \nu X)=\frac{G_F^2m_c^5}{192\pi^3}|V_{cq}|^2\cdot \{
(z_0(x)-\frac{2\al_s(m_c^2)}{3\pi}(\pi^2-25/4) z_0^{(1)}(x))\cdot$$
$$\cdot (1-K_D/m_c^2+\frac{1}{4}G_D/m_c^2) -
z_1(x) G_D/m_c^2 +{\cal O}(\al_s^2, \al_s/m_c^2, 1/m_c^3)\} \eqno(16)$$
where the phase space
factors $z$ account for the mass of the quark $q=s,d$ in the final
state:
$$z_0(x)=1-8x+8x^3-x^4-12x^2\log{x} \;\;,\;\; z_1(x)=(1-x)^4\;\;\;, $$
$$z_0^{(1)}(0)=1\;\;,\;\;  z_0^{(1)}(1)= 3/(2\pi^2-25/2))\simeq 0.41
\;\;\;,\;\;\;x=(m_q/m_c)^2 \eqno(17a)$$
(the function $z_0^{(1)}$ can be found in ref.\cite{rc}),
while $K_D$ and $G_D$ denote
the kinetic energy and the chromomagnetic matrix elements respectively:

$$G= \frac{1}{2M_D}\matel{D}{\bar{c} \sG c}{D} \simeq
\frac{3}{2}\cdot(M_{D^*}^2-M_D^2) \;\;,\;\;\;K = \frac{1}{2M_D}
\matel{D}{\bar{c}\frac{(i\ve D)^2}{2} c}{D})\equiv \frac{\pp}{2}
\;\;.\eqno(17b)$$
It should be noted that the explicit form of the order $\al_s$
perturbative correction in
eq.(16) refers to the on shell (pole) definition for the charm quark mass.

The expressions given above allow one to determine the mass of the $c$ quark
from the measured semileptonic width, provided the weak
mixing angles $|V_{cs}|$ and $|V_{cd}|$ are known. We shall assume in our
subsequent analysis that the weak mixing matrix is determined by the
existence of only three generations; the quantities $|V_{cs}|$ and
$|V_{cd}|$ are then known with an accuracy that is more than sufficient for
our purposes\footnote{There are some direct experimental measurements of
these two angles, however they suffer from relatively large uncertainties
for $V_{cs}$.}. Adopting for the semileptonic widths of $D$ mesons the value
of $\Gamma\ind{SL}=\BR\ind{SL}(D^+)\cdot \tau_{D^+}^{-1} \simeq
1.06\cdot10^{-13}\GeV$ and
assuming $\pp \simeq .3 \GeV^2$ we then find
$$m_c=(1.57 \pm 0.03)\GeV\;\;$$
where we have included only the
experimental error, coming mainly from $\Gamma\ind{SL}$. We used here the
values $\al_s(m_c^2)=.33$ and $m_s=140\MeV$.

The smallness of the error in $m_c$ quoted above reflects
the fact that the width
depends on the fifth power of $m_c$.
The value found for $m_c$ is then also not
very sensitive to details in the nonperturbative corrections to the
spectator picture; these theoretical uncertainties will be discussed next.

The perturbative ${\cal O}(\al_s)$ term
reduces the width in eq.(16) by about $25\%$; the leading
nonperturbative corrections $\sim {\cal O}(1/m_c^2)$
are of comparable size: the chromomagnetic term
and the kinetic term yield a reduction by
$\sim 25\%$ and by about $6\%$, respectively.
However all these effects do not generate a prominent change in the
value of $m_c$:  the perturbative corrections increase $m_c$
by $75\MeV$;
the chromomagnetic term and the kinetic energy term force
$m_c$ up by $85\MeV$ and by
$20\MeV$, respectively, for the stated value of $\pp$.
 From these numbers we infer that the associated uncertainties in these
corrections are rather insignificant.
There is some uncertainty concerning the value of $\alpha _S(M_Z^2)$
and the scale at which to evaluate $\alpha _S$ in charm decays; yet those
effects are quite unlikely to exceed 20\% and can be addressed by
including ${\cal O} (\al_s^2)$ contributions.
Potentially
larger errors can be expected from the nonperturbative effects. As discussed
in the previous section the uncertainty in the chromomagnetic field could
conceivably be of order $30\%$ in the charm system. Corrections of similar
size can be expected from higher power terms in the $1/m_c$ expansion.
Finally the exact value of the kinetic term is not known. Yet
its impact is generally somewhat suppressed as compared to the
chromomagnetic interaction, and is dominated by the latter for reasonable
sizes \cite{VLB} of the mean Fermi momentum of the heavy quark; the
dependence on the $\pp$ is illustrated in Table~1. It is worth adding that
for obvious reasons there is no significant dependence of the extracted
value for $m_c$ on $m_s$ when the latter is varied within reasonable limits
for a current quark mass.

Combining all of this we then estimate the
present theoretical uncertainty in extracting $m_c$ to be about $30\MeV$;
to be conservative one may increase it up to say $50\MeV$:
$$ m_c=(1.57 \pm 0.03\pm 0.05)\GeV\; .\eqno(18)$$
A more detailed
understanding of the intrinsic limitations on the accuracy of such approach
can be expected~\cite{BUV} in the future from explicit calculations of the
higher order corrections, see e.g. refs.~\cite{BSV,BSUV2}.

Quite often another definition of the quark mass is used, namely the $\MS$
one corresponding to a Euclidean renormalization point $q^2 = - m_Q^2$.
At the one-loop level they are related by

$$ m_Q^{\MS}(-m_Q^2) \simeq  m_Q^{\pole}\cdot
(1-\frac{4}{3}\frac{\al_s(m_Q^2)}{\pi})\;\;;\eqno(19)$$
eq.(18) then yields $1.35\GeV$ for this definition of $m_c$.

It might appear at this point
that the fifth power dependence of the decay width on the
mass of the quark allows one to make a rather accurate extraction of $m_c$
almost without any detailed information about the nature of nonperturbative
corrections.  Such a conclusion would however overstate the
facts: for its validity rests on the  absence of
nonperturbative corrections of order $1/m_Q$
in the total widths, as proven in
refs.~\cite{BUV,BS}. This explained {\em a posteriori} why simple
minded estimates made a long time ago that ignored nonperturbative
corrections yielded a charm quark mass of around $1.5\GeV$.
The real
shortcoming of these estimates was the following more subtle point:
in these models
one cannot distinguish between the mass of the charm {\em quark} and
of the charm {\em hadron} in an unambigous fashion; not surprisingly
the estimates numerically fell
somewhere in between. This alternative of course is a reformulation of the
problem of nonperturbative $1/m_Q$ corrections to widths. The QCD approach
ensures that if $m_c$ is understood as the (current) quark mass then these
corrections are absent! This is a consequence of the conservation of
the colour flow in QCD as can be seen by simple quantum mechanical
arguments (see ref.~\cite{BSUV2} for details).

Having extracted a value for the charm quark mass one can then
determine the mass of the beauty quark by employing an expansion of the
heavy flavour hadron
masses in terms of the heavy quark mass \cite{BUV1}:
$$ m_b-m_c\,= \, \frac{M_{B}+3M_{B^*}}{4} - \frac{M_{D}+3M_{D^*}}{4} \,+ \,
\pp \cdot (\frac{1}{2m_c}-\frac{1}{2m_b}) \,+\, {\cal O}(1/m_c^3\,,\;
1/m_b^3) \;; \eqno(20)$$
unfortunately the accuracy of this expansion is
obviously controlled by the charm quark mass. Note that the absence of the
perturbative corrections on the left-hand side of eq.(20)
implies that pole masses are
assumed. For the same value of the kinetic term the central value appears to
be $5.0\GeV$ and $4.5\GeV$ for the pole and the $\MS$ masses of the $b$
quark, respectively. For such indirectly determined mass of the beauty quark
the dependence on the size of $\pp$ is more significant as is illustrated by
Table~1.

The same consideration fixes also the value of the scale $\bar\Lambda$ that
determines the asymptotic mass difference between the mass of the heavy
flavour hadron and the mass of the constituent heavy quark. For the family
of the lowest lying pseudoscalar and vector mesons one then has
$\bar\Lambda\approx 300\MeV$.
This value is also correlated with the size of
the kinetic term, and the latter seems to represent now the main uncertainty
in the value of $\bar\Lambda$. Note that we have shown in a recent paper
\cite{BUlam} that there is no lower bound on $\La$
in contrast to a recent claim
\cite{MANOHAR}; {\em a priori} $\La$ could have been even negative (see
also \cite{BSUV2}).

Above we have
discussed only the most obvious uncertainties that one encounters in the
numerical evaluation of the heavy quark mass.
There is a number of additional
purely theoretical corrections that in principle can affect them. Those
corrections were
discussed in some detail in a recent paper~\cite{BSV}; a more
comprehensive analysis will be given in
a forthcoming publication~\cite{BSUV2}. We therefore do not
dwell on them here and only comment that they are not expected
to affect significantly the values extracted for the masses.
\vspace*{.2cm}

 Turning to the extraction of the quark
mixing parameter $|V_{cb}|$ we first note
that phenomenological studies have shown that the total widths and lepton
spectra in semileptonic B decays depend mostly on the difference $m_b-m_c$
rather than on the absolute values of the heavy quark masses. In the
framework of the heavy quark expansion these findings are the reflection of
the fact that the $c$ quark is rather heavy
even as seen from the scale of the
$b$ quark mass; they can be understood as meaning that the error one makes
in using the measured masses of the charm and beauty {\em hadrons} rather
than those of the charm and beauty {\em quarks} is reduced relative to its
natural scale $1/m_Q$. This error can actually be reduced even further by
using values for the quark masses as they were extracted above
from the widths of $D$ mesons. A value for $|V_{cb}|$ can then be
obtained from the measured semileptonic width.

 From a theoretical point of view such an inclusive method has some clear
advantages over an extraction of $|V_{cb}|$ from the exclusive decays $B\ra
D,D^* l\nu$ suggested by the Heavy Quark Symmetry -- even when taken at the
``gold plated'' point of zero recoil \cite{MN}. For the accuracy
of the symmetry is governed by the {\em lightest} quark mass in the process,
namely  $1/m_c^2$; on the other hand the expansion parameter for the
inclusive widths is an inverse power of $m_b$, and this method can be
applied even for light quarks in the final state.

The necessary expression for the width is given by eq.(16) where
the obvious substitutions $b$ for $c$ and $c$ for $s$
are made. Using the values $\tau_B \simeq 1.4\,\mbox{ps}$ and
$\BR\ind{SL}(B)\simeq .105$~\footnote{For the sake of definiteness we used
$\Gamma\ind{sl}(b\ra u)=.01 \Gamma\ind{sl}(b\ra c)$.} we arrive at the
results shown in Table~1. The main dependence is again due to the kinetic
term, however now it is rather weak; the uncertainty associated with the
value of $\al_s$ is now smaller owing to the higher energy scale. It is
tempting to conclude then that this method provides us at least three times
better accuracy for the extraction of $|V_{cb}|$ than other methods
discussed so far.

The values for $m_c$ and $m_b$ that we have obtained in this Section
are around the upper end of existing estimates.
More conventional values emerge if one uses a smaller kinetic energy
term of about $(.4\GeV)^2$. It is worth noting that the theoretical
predictions of the QCD sum rules for $\bar\Lambda$ prefer smaller values of
about $400\MeV$ \cite{VLB2}; using this smaller value of $\bar\Lambda$
decreases the predicted value of the kinetic term~\cite{VLB}. On the other
hand it is most natural to expect~\cite{BSUV2} the mean value of the Fermi
momentum not to exceed significantly $\bar\Lambda$. Therefore we think that
values for the $b$ quark mass around $4.95\GeV$ together with a
relatively small scale for the Fermi momenta of the order of $300 \div
500\MeV$ are both selfconsistent and acceptable phenomenologically. This
hypothesis can and will be cross-checked in detailed study of spectra
in semileptonic and radiative $B$ decays.

% \section{Extraction of $|V_{cb}|$}

\section{Conclusion}

In this note we have applied a systematic expansion in $1/m_Q$ that exists
for the inclusive widths, to a few phenomenologically interesting issues
concerning the properties of charm and beauty mesons. The general method
we use has been suggested earlier \cite{BUV,BS,VSlog} and refined in
subsequent papers~\cite{BBSUV,BSUV,BU2,BSV,BSUV2}. The main object of our
analysis
was the pattern of the charm meson widths, including the $D_s$ width.

Generally the size of preasymptotic nonperturbative effects is almost of
order unity in charm; this makes it difficult to arrive at
conclusions that are both definite and reliable. The
detailed classification of the corrections enable us to conclude
nevertheless that the width splitting between $D_s$ and $D^0$ must be
reasonably small. There are four sources for this difference that can
produce effects on the few percent level:  1) A larger Lorentz
reduction of
the decay probability for the $c$ quark inside the $D_s$
meson due to a more rapid Fermi motion. 2) Destructive
interference in the Cabibbo suppressed decays of $D_s$; each of these
effects are estimated to decrease the width of $D_s$ by $3\div5\%$. 3) The
mode $D_s\ra \tau \nu_{\tau}$ increases it by about $3\%$.
4) Nonleptonic decays of
both mesons are affected by WA which is present also in semileptonic decays
of $D_s$. According to the analysis of ref.~\cite{BU2} all hadronic
parameters governing the strength of WA in different decays can be obtained
by a careful comparison of the lepton spectra in semileptonic decays of
charged and neutral $B$ mesons. However such data do not
exist for now; therefore the size of WA remains an
unknown and -- unfortunately
-- the potentially largest effect numerically in
$\tau (D_s)/\tau (D^0)$.

Other lines of reasoning suggest that the impact of WA on
inclusive charm decays can hardly be large: presumably
it does not exceed
$10\div20\%$ of the total $D^0$ width;
furthermore it can be of either sign! We
see therefore that the data on the lifetimes of charm mesons
fit reasonably well the theoretically expected pattern. On the other hand
the existing experimental determination of the $D_s$ lifetime can be viewed
as some phenomenological constraint on the non-factorizable matrix elements of
the four-fermion operators in charm mesons:

$$|\frac{f^2_{D_s}}{f^2_D}(g^{(D_s)}\ind{singl}-0.25 g^{(D_s)}\ind{oct})
+ 0.03g^{(D)}\ind{singl}-0.3g^{(D)}\ind{oct} |\; \lsim \;
0.01\cdot(\frac{200\MeV}{f_D^2})^2 \eqno(21)$$
where we have used the notations introduced in eq.(16) of ref.~\cite{BU2}.
Of course the extremely small number in the {\em rhs} of eq.(21) should not
be taken too literally.

The theoretical prediction for the $D_s$ lifetime can be further
clarified by
accurate measurements of the masses of $B_s$ and $B_s^*$ masses on the one
hand and
on the other a better understanding of the scale of Fermi motion in heavy
hadrons which can be
obtained from an accurate analysis of spectra in semileptonic and/or
radiative decays of beauty particles.

Similar preasymptotic corrections splitting the widths of beauty particles
are expected to be essentially smaller~\cite{BU}. A careful analysis of
the QCD corrections \cite{VSKU,BU} has lead to the observation that the
width difference between the two mass eigenstates in the $B_s$-$\bar B_s$
sector quite probably represents the largest numerical difference in the
family of beauty mesons:

$$\frac{\Delta\Gamma_{B_s}}{\bar\Gamma} \simeq .15 \frac{f_{B_s}^2}
{(200\MeV)^2}
\eqno(22)$$
This estimate is valid for $f_{B_s}$ -- which acts as an expansion
parameter -- not too large. To a very good approximation one can identify
the two mass eigenstates as \CP eigenstates. Obviously the upper bound for
the width difference is reached when all final states in the decay channel
$b\ra c \bar c s$ for the decays of $B_s$ have the same \CP parity and
$|\Delta\Gamma_{B_s}| \simeq 2\Gamma(b\ra c \bar c s)$. The rough estimate
for this partial decay width is given by the parton expression and
constitutes about $20\%$, therefore the estimate eq.(22) is sensible up to
$f_{B_s}\lsim 300\MeV$.

A straightforward, in principle, application
of a $1/m_Q$ expansion of
inclusive decay widths~\cite{BUV,BS} is the determination of the mass
of charm quark from the semileptonic $D$ width, then obtaining
the $b$ quark mass from the mass formulae and
subsequently extracting $|V_{cb}|$
from the $B$ meson semileptonic width.
In charm non-perturbative effects
dominate the more familiar perturbative corrections:
the former constitute about $35\%$
whereas ${\cal O}(\al_s)$ corrections yield approximately $25\%$. None of
these however produce a significant change in the resulting value of $m_c$
owing to the fifth power dependence of the decay width on the mass. In fact
this conclusion is a consequence of the non-trivial fact
that there are no corrections of order
$1/m_c$ to the heavy flavour widths, as shown
in refs.~\cite{BUV,BS}; this is a peculiar feature of the gauge
nature of the strong interaction producing the bound state in the initial
state and driving hadronization dynamics in the final states, and it
reflects the conservation of the colour current. A reasonable estimate for
the theoretical
uncertainty of the value of the $c$ quark is about $50\MeV$ and one can
count on it being decreased in the future.

The masses of $b$ and $c$ quarks obtained in such a way naively seem to be
some $100\MeV$ higher than the conventional values inferred from the direct
QCD analysis, and in particular from estimates
based on charmonium sum rules. In
fact a careful analysis\footnote{It was elaborated in joint
discussions with M.Shifman and A.Vainshtein; we are grateful for their
permission to mention it prior to publication of that result.}
undertaken recently suggested a rational explanation of this difference in
the framework of QCD. We will report on it in the forthcoming
paper~\cite{BSUV2}.

Using the observed semileptonic width of $B$ mesons and including
the leading corrections to the parton formulae one obtains the value of the
KM mixing angle defining the strength of the $b \ra c$ transitions:

$$|V_{cb}|\simeq 0.043 \;\;\;.$$
This method seems to be the most accurate and realiable way to obtain the
value of $|V_{cb}|$. For its accuracy is governed by powers of $1/m_b$
as confronted to the decays into exclusive final state $D$ and $D^*$
\cite{MN} where
actually the $c$ quark sets the mass scale for the corrections to the Heavy
Spin-Flavour Symmetry. The corrections to the inclusive decays have been
calculated explicitely through order $1/m_b^3$ and in principle the
expansion can be extended, whereas $1/m_c^2$ effects for the
exclusive decays \cite{FN} may already constitute the limiting factor
for improving the estimates.

Our expressions for the semileptonic decay widths can be easily translated
into the semi-phenomenological parameters $z_c$ and $z_u$ that were
introduced long ago to account for the final quark mass suppression in
inclusive semileptonic widths as well as for all other possible corrections
to the parton formulae. Their usual definition was

$$\Gamma\ind{SL}(b\ra q) = z_q \cdot
\frac{G_F^2 (5\GeV)^5}{192\pi^3}|V_{qb}|^2\;\;\;,\;\;q\,=\,u,c\;\;\;.
\eqno(23)$$
For the thus defined factor $z_c$ we get values varying from $.36$ at small
$\pp$ to $.43$ for $\pp \simeq 0.8\GeV^2$; these values seem to agree with
the estimates obatined from experiment using fits based on some
phenomenological models~\cite{ACM} of a heavy quark decay. The ratio
$z_u/z_c \simeq 2.08$ appears to be independent of the scale of the Fermi
motion, which also reproduces the expectations that
one inferred from those models.

The
phenomenological extraction of $m_c$ and $m_b$ enables one to determine the
hadronic parameter $\bar\Lambda$ which in the nonrelativistic description of
a heavy hadron plays a role of the constituent mass of the spectator(s). As
was stated in ref.~\cite{BU2} $\bar\Lambda$ actually completely defines the
leading, $1/m_Q$, non-perturbative
shift in the average invariant mass of the
final hadrons in semileptonic or radiative decays. In the quasi two
particle decays like $b\ra s+\ga$ the correction is given by
$\;\de m^2\ind{np}\simeq \La m_b +{\cal O}(\hscale^2)$ if one
neglects the mass of strange quark. We see that the average final
state hadronic
mass is increased by some $1.5\div2\GeV^2$ as compared to a
perturbative estimate.  We note that once again the leading nonperturbative
effects seem to be at least comparable in size to the perturbative
corrections~\cite{AG} even in beauty decays. Accordingly they have to be
included in a proper quantitative treatment; together with the effect of the
Fermi motion the $\La$ parameter describes the significant $1/m_b$
corrections to the spectrum of photons, which in turn define the relative
weight low lying exclusive final states can command in such decays.

%***
\vspace*{1cm}

{\it Note added}: While this paper was written up we become
aware \cite{MANNEL} of the work of ref.\cite{LUKE} which have a significant
overlap with the part concerning the determination of quark masses and the
mixing parameter. At that point we
present a more detailed discussion of nonperturbative corrections
as they are relevant for extracting $m_c$, $m_b$ and $|V(cb)|$.
The two treatments thus complement each other. As was mentioned in Section 3
we also believe that using the upper bound of Guralnik and Manohar -- that
constitutes an important element of the analysis of paper \cite{LUKE} -- is
irrelevent \cite{BUlam} in this context.

\vspace*{1cm}

{\bf ACKNOWLEDGEMENTS:} \hspace{.4em} The impetus for this analysis
and several of the main ideas came from illuminating discussions one
of us (N.U.) had with J.Rosner during a stay at the University of
Chicago. We gratefully acknowledge promotional joint studies and
discussions of theoretical aspects of the problems addressed in the present
paper with A.Vainshtein and M.Shifman; N.U. also thanks V.Yu.Petrov
for informative
conversations on the subject of this paper and
V.Braun for discussion of the QCD sum rules for heavy quarks.
This work was supported by the National Science Foundation under
grant number PHY 92-13313.

\vspace*{3cm}

\begin{flushleft}
\begin{tabular}{|c|c|c|c|c|c|c|}\hline
$\frac{\matel{D}{\bar{c}(i\ve D)^2 c}{D}}{2M_D}$ & $m_c,\GeV$
& $m_c\,(\MS),\GeV$ & $m_b,\GeV$ & $m_b\,(\MS),\GeV$ &
$\La,\GeV$ & $|V_{cb}|$ \\ \hline
$(0.1\GeV)^2$ & 1.55 & 1.33 & 4.89 & 4.43 & 0.422 & 0.0442 \\
$(0.3\GeV)^2$ & 1.56 & 1.34 & 4.91 & 4.45 & 0.390 & 0.0437 \\
$(0.5\GeV)^2$ & 1.57 & 1.35 & 4.96 & 4.50 & 0.327 & 0.0427 \\
$(0.7\GeV)^2$ & 1.59 & 1.36 & 5.03 & 4.56 & 0.234 & 0.0413 \\
$(0.8\GeV)^2$ & 1.60 & 1.37 & 5.07 & 4.60 & 0.177 & 0.0404 \\
$(0.9\GeV)^2$ & 1.61 & 1.39 & 5.12 & 4.64 & 0.112 & 0.0395 \\ \hline
\end{tabular}
\end{flushleft}
{\Large{Table 1:}} Dependence of the extracted parameters on the size of
the kinetic energy operator in nonstrange mesons. We used here the strength
of the QCD running coupling corresponding to $\Lambda\ind{QCD}=180\MeV$.

\end{document}